\documentclass[12pt]{article}

\usepackage[T1]{fontenc}
\usepackage[utf8]{inputenc}
\usepackage{lmodern}
\usepackage[a4paper,margin=2.45cm]{geometry}
\usepackage{amsmath,amssymb,mathtools,bm}
\usepackage{authblk}
\usepackage{booktabs}
\usepackage{microtype}
\usepackage{xcolor}
\usepackage{pgfplots}
\usepackage{cite}
\usepackage{lineno}
\usepackage{hyperref}

\pgfplotsset{compat=1.18}
\hypersetup{
  colorlinks=true,
  linkcolor=blue!55!black,
  citecolor=blue!55!black,
  urlcolor=blue!55!black
}

\newcommand{\dd}{\mathrm{d}}
\newcommand{\ee}{\mathrm{e}}
\newcommand{\ii}{\mathrm{i}}
\newcommand{\Tr}{\operatorname{Tr}}
\newcommand{\mpl}{\ell_{\mathrm{Pl}}}

\title{Canonical-Ensemble Stability of the Quantum
Oppenheimer--Snyder Black-Hole Exterior}
\author{Wen-Xiang Chen\\
\small School of Electronic Information\\ Guangzhou City University of Technology, Guangzhou 510800, China\\
\small E-mail: wxchen4277@qq.com}
\date{}

\begin{document}

\maketitle

\begin{abstract}
We study the local canonical thermodynamics of the quantum
Oppenheimer--Snyder (qOS) black-hole exterior previously obtained from loop
quantum cosmology. The spacetime is adopted from the literature; no new
black-hole solution, regular extension, or singularity-resolution mechanism
is claimed here. At fixed quantum parameter $\alpha$, we rewrite the outer
horizon in a dimensionless form and obtain exact expressions for the mass,
Hawking temperature, entropy, and heat capacity. These expressions reproduce
the logarithmic entropy term and the heat-capacity divergence reported in
earlier analyses. We then place those results in an explicit, regulated
canonical-ensemble framework. The off-shell canonical action has a local
minimum on a near-extremal branch with positive heat capacity and a local
maximum on a large-black-hole branch with negative heat capacity. The two
saddles merge at
$r_h/\sqrt{\alpha}=\sqrt{(4+2\sqrt{7})/3}$, where the temperature is maximal
and the heat capacity diverges. Because the asymptotically flat partition
function is not normalizable without boundary data, this divergence is
interpreted as a Davies-type local stability transition, not by itself as a
global first-order phase transition. The principal contribution of this
work is therefore a transparent canonical-saddle organization of known qOS
thermodynamic results, together with a precise statement of its domain and
limitations.
\end{abstract}

\noindent\textbf{Keywords:} loop quantum gravity; quantum
Oppenheimer--Snyder model; black-hole thermodynamics; canonical ensemble;
heat capacity

\section{Introduction}
\label{sec:intro}

Black-hole thermodynamics connects horizon geometry, quantum field theory,
and statistical mechanics through the Bekenstein--Hawking entropy and Hawking
temperature \cite{Bekenstein1973,Hawking1975}. A Euclidean gravitational path
integral gives a saddle-point realization of a thermal partition function
\cite{GibbonsHawking1977}, while the analysis of a black hole in a finite
cavity makes clear that boundary conditions are essential for a well-defined
canonical ensemble \cite{York1986}.

Loop quantum gravity (LQG) provides a background-independent quantization of
geometry \cite{Rovelli2004,AshtekarLewandowski2004}. In a symmetry-reduced
collapse model, Lewandowski \emph{et al.}\ derived a quantum-corrected
Oppenheimer--Snyder exterior by matching an effective loop-quantum-cosmology
interior to a static exterior \cite{LewandowskiEtAl2023}. The present
manuscript adopts that exterior geometry. It neither derives the metric nor
constructs a new regular black-hole solution.

The qOS geometry has motivated studies of thermodynamics, wave propagation,
and higher-dimensional or cosmological-constant extensions
\cite{Zhang2023,LinZhang2024,ShiZhangMa2024,OuZhang2025,JiangLinZhang2026}.
Related loop-quantum black-hole geometries have also been explored through
gravitational lensing \cite{LiZhang2024}, strong cosmic censorship
\cite{LinZhangBravo2025}, spinning-particle dynamics
\cite{DuLiuZhang2025}, particle deflection \cite{LiZhang2025}, and
island-based information diagnostics \cite{DuSunZhang2026}. Recent preprints
consider primordial-black-hole phenomenology and evaporation
\cite{WangZhangPBH2026,WangZhangEvap2026}. These works form the necessary
context for the effective metric and prevent established results from being
mistaken for results original to this article.

Our scope is deliberately narrow. First, we state the LQG kinematical input
without conflating an area spectrum with the complete operator definition.
Second, we record the qOS scalar equation and thermodynamic formulae with
explicit attribution. Third, we formulate the canonical ensemble and use an
off-shell action to classify its local saddles. The exact dimensionless
parametrization makes the heat-capacity pole and the two-saddle structure
especially transparent. We do not evaluate a regulated one-loop determinant,
specify a complete quantum interior, or claim a global phase diagram.

\section{Kinematical input and the effective exterior}
\label{sec:geometry}

\subsection{What the area spectrum does and does not define}

For a spin-network state $|\Gamma,\{j_e\},\{\iota_v\}\rangle$ and a surface
$S$ intersecting graph edges transversely away from vertices, the standard
area operator has the eigenvalue action
\begin{equation}
 \widehat{A}(S)|\Gamma,\{j_e\},\{\iota_v\}\rangle
 =
 8\pi\gamma\mpl^2
 \sum_{p\in S\cap\Gamma}\sqrt{j_p(j_p+1)}
 |\Gamma,\{j_e\},\{\iota_v\}\rangle .
 \label{eq:area-spectrum}
\end{equation}
Here $\gamma$ is the Barbero--Immirzi parameter and $j_p$ labels the
representation carried by the edge crossing $S$. Equation
\eqref{eq:area-spectrum} is the spectrum on this class of states, not the full
definition of $\widehat{A}(S)$. The operator itself is obtained by
regularizing the classical area functional in terms of flux operators; more
general intersections require the corresponding recoupling rules
\cite{AshtekarLewandowskiArea1997,AshtekarLewandowski2004}.

The volume operator is not needed in the calculation below, so no schematic
volume spectrum is introduced. Unlike Eq.~\eqref{eq:area-spectrum}, its
action depends on vertices, edge orientations, intertwiners, and the chosen
regularization \cite{AshtekarLewandowskiVolume1998}. Removing an
oversimplified volume formula avoids attributing to it a universal
single-spin spectrum.

\subsection{Adopted qOS exterior and its domain}

In units $c=\hbar=k_{\mathrm B}=1$, the exterior line element derived in
Ref.~\cite{LewandowskiEtAl2023} is
\begin{align}
 \dd s^2&=-f(r)\,\dd t^2+\frac{\dd r^2}{f(r)}
          +r^2\dd\Omega_2^2,                                      \label{eq:metric}\\
 f(r)&=1-\frac{2GM}{r}+\frac{\alpha G^2M^2}{r^4},                  \label{eq:lapse}\\
 \alpha&=16\sqrt{3}\,\pi\gamma^3\mpl^2 .                           \label{eq:alpha}
\end{align}
Thus $\alpha$ has dimensions of length squared; it is not dimensionless.
Writing $m=GM$, the matching construction fixes the exterior only in its
physical domain, conventionally bounded by the bounce scale
$r\geq r_b=(\alpha m/2)^{1/3}$ \cite{LewandowskiEtAl2023,Zhang2023}. A
formal extrapolation of Eqs.~\eqref{eq:metric}--\eqref{eq:lapse} to $r=0$ is
not justified by that construction. Consequently, regularity of a complete
spacetime cannot be inferred from the exterior lapse alone.

Let $r_h$ be a horizon and set
\begin{equation}
 x=\frac{r_h}{\sqrt{\alpha}},\qquad
 q=\sqrt{x^2-1},\qquad
 \mu=\frac{m}{\sqrt{\alpha}}=\frac{GM}{\sqrt{\alpha}} .
 \label{eq:dimensionless}
\end{equation}
Solving $f(r_h)=0$ on the branch that approaches Schwarzschild as
$\alpha\to0$ gives
\begin{equation}
 m(r_h)=\frac{r_h^2}{r_h+\sqrt{r_h^2-\alpha}},
 \qquad
 \mu(x)=\frac{x^2}{x+q}.                                          \label{eq:mass}
\end{equation}
The minimum mass and degenerate horizon occur at
\begin{equation}
 x_e=\frac{2}{\sqrt{3}},\qquad
 r_e=2\sqrt{\frac{\alpha}{3}},\qquad
 M_e=\frac{4\sqrt{\alpha}}{3\sqrt{3}\,G}.                         \label{eq:extremal}
\end{equation}
Only the outer, Schwarzschild-connected branch $x\geq x_e$ is used below.

\section{Scalar probe: statement and attribution}
\label{sec:scalar}

A minimally coupled scalar of mass $\mu_s$ obeys
\begin{equation}
 \left(\Box-\mu_s^2\right)\Phi=0.                                 \label{eq:KG}
\end{equation}
With
\begin{equation}
 \Phi(t,r,\vartheta,\varphi)
 =\frac{\psi_{\ell\omega}(r)}{r}
   Y_{\ell m}(\vartheta,\varphi)\ee^{-\ii\omega t},
 \qquad
 \frac{\dd r_*}{\dd r}=\frac{1}{f(r)},
\end{equation}
Eq.~\eqref{eq:KG} becomes
\begin{equation}
 \frac{\dd^2\psi_{\ell\omega}}{\dd r_*^2}
 +\left[\omega^2-V_\ell(r)\right]\psi_{\ell\omega}=0,
 \qquad
 V_\ell(r)=f(r)\left[
 \frac{\ell(\ell+1)}{r^2}+\frac{f'(r)}{r}+\mu_s^2
 \right].
 \label{eq:radial}
\end{equation}
Equations of this type, including perturbative and quasinormal-mode analyses
for qOS-related metrics, have already been studied in
Refs.~\cite{LinZhang2024,ShiZhangMa2024,OuZhang2025}. They are included only
to fix notation and are not counted as an original result of this work. No
new spectrum is computed here.

\section{Exact thermodynamic relations at fixed \texorpdfstring{$\alpha$}{alpha}}
\label{sec:thermo}

\subsection{Temperature and entropy}

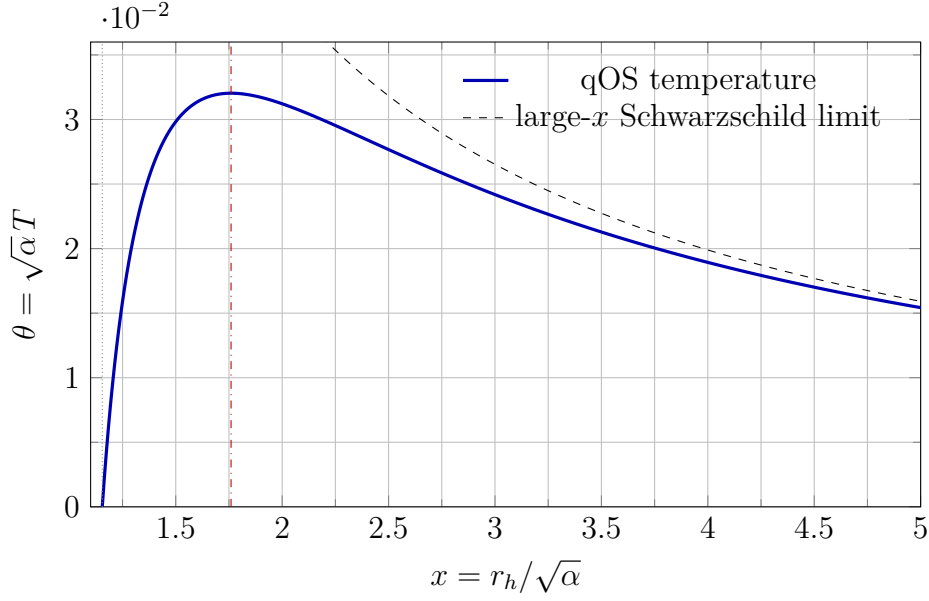
\begin{figure}[htbp]
 \centering
 \begin{tikzpicture}
  \begin{axis}[
   width=0.78\textwidth,
   height=0.48\textwidth,
   xlabel={$x=r_h/\sqrt{\alpha}$},
   ylabel={$\theta=\sqrt{\alpha}\,T$},
   xmin=1.1,xmax=5,
   ymin=0,ymax=0.036,
   grid=both,
   minor tick num=1,
   legend style={draw=none,fill=none},
   legend pos=north east
  ]
   \addplot[blue!70!black,very thick,domain=1.154701:5,samples=400]
   {(2*sqrt(x^2-1)-x)/(2*pi*x*(x+sqrt(x^2-1)))};
   \addlegendentry{qOS temperature}
   \addplot[black,dashed,domain=1.154701:5,samples=200]
   {1/(4*pi*x)};
   \addlegendentry{large-$x$ Schwarzschild limit}
   \addplot[gray,densely dotted] coordinates
   {(1.1547005,0) (1.1547005,0.036)};
   \addplot[red!70!black,dashdotted] coordinates
   {(1.7598771,0) (1.7598771,0.036)};
  \end{axis}
 \end{tikzpicture}
 \caption{Exact dimensionless Hawking temperature. The dotted line marks the
 extremal point $x_e=2/\sqrt{3}$ and the dash-dotted line marks the maximum
 at $x_c=\sqrt{(4+2\sqrt7)/3}$.}
 \label{fig:temperature}
\end{figure}

The surface-gravity temperature of the outer horizon is
\begin{align}
 T(r_h)
 &=\frac{f'(r_h)}{4\pi}
 =\frac{2\sqrt{r_h^2-\alpha}-r_h}
 {2\pi r_h\left(r_h+\sqrt{r_h^2-\alpha}\right)},                 \label{eq:temperature}\\
 \theta(x)
 &\equiv\sqrt{\alpha}\,T
 =\frac{2q-x}{2\pi x(x+q)} .                                     \label{eq:theta}
\end{align}
It vanishes at $x_e$ and tends to the Schwarzschild result at large $x$.
The exact curve is shown in Fig.~\ref{fig:temperature}.

Keeping $\alpha$ fixed, the first law $\dd M=T\,\dd S$ gives
\begin{equation}
 S(r_h)=\frac{\pi}{G}\left[
 r_h\sqrt{r_h^2-\alpha}
 +\alpha\ln\!\left(
 \frac{r_h+\sqrt{r_h^2-\alpha}}{\sqrt{\alpha}}
 \right)\right]+S_0 .                                            \label{eq:entropy}
\end{equation}
We choose the additive constant so that the extremal entropy vanishes,
\begin{equation}
 S_0=-\frac{\pi\alpha}{G}
 \left(\frac{2}{3}+\frac{1}{2}\ln3\right).                       \label{eq:S0}
\end{equation}
This convention affects absolute free energies but not equilibrium or local
stability. In terms of the horizon area $A=4\pi r_h^2$, the large-area
expansion is
\begin{equation}
 S=\frac{A}{4G}
 +\frac{\pi\alpha}{2G}\ln\!\left(\frac{A}{\pi\alpha}\right)
 +\text{constant}
 +\mathcal{O}\!\left(\frac{\alpha^2}{GA}\right).                  \label{eq:entropy-large}
\end{equation}
Equation~\eqref{eq:entropy}, rather than an ad hoc fluctuation ansatz, is the
entropy used throughout this manuscript. Related qOS thermodynamics and
logarithmic terms have already been derived in
Refs.~\cite{LinZhang2024,ShiZhangMa2024,OuZhang2025,JiangLinZhang2026}; the
present calculation is a compact reparametrization and consistency check.

\subsection{Heat capacity and the stability-transition feature}

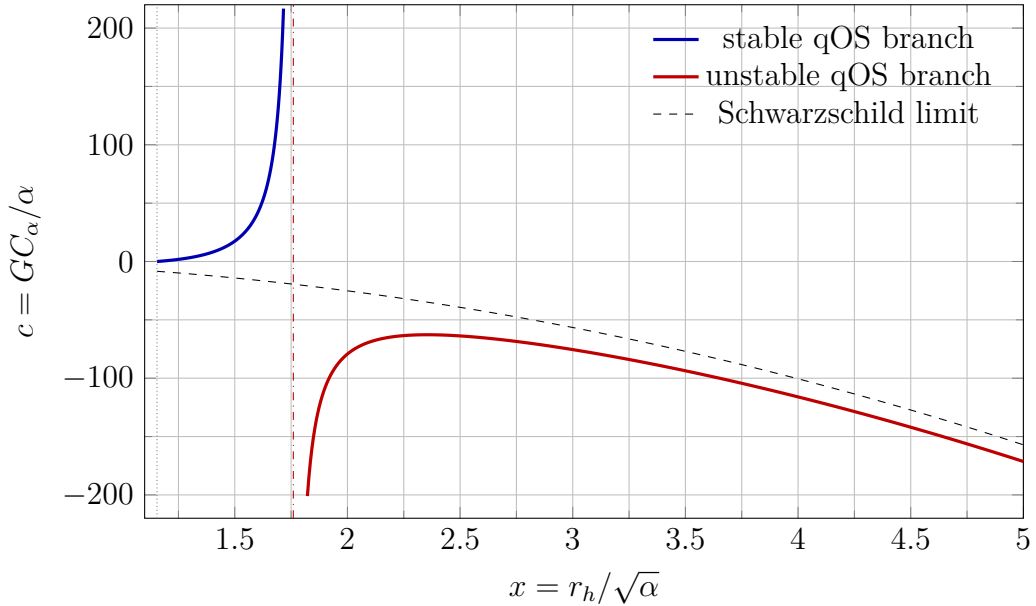
\begin{figure}[htbp]
 \centering
 \begin{tikzpicture}
  \begin{axis}[
   width=0.82\textwidth,
   height=0.52\textwidth,
   xlabel={$x=r_h/\sqrt{\alpha}$},
   ylabel={$c=GC_\alpha/\alpha$},
   xmin=1.1,xmax=5,
   ymin=-220,ymax=220,
   grid=both,
   minor tick num=1,
   unbounded coords=jump,
   legend style={draw=none,fill=none,at={(0.98,0.98)},anchor=north east}
  ]
   \addplot[blue!70!black,very thick,domain=1.154701:1.756,samples=350,
            restrict y to domain=-220:220]
   {2*pi*x^3*(x*sqrt(x^2-1)+x^2-2)/
    (x*(4-x^2)+sqrt(x^2-1)*(2-x^2))};
   \addlegendentry{stable qOS branch}
   \addplot[red!75!black,very thick,domain=1.764:5,samples=500,
            restrict y to domain=-220:220]
   {2*pi*x^3*(x*sqrt(x^2-1)+x^2-2)/
    (x*(4-x^2)+sqrt(x^2-1)*(2-x^2))};
   \addlegendentry{unstable qOS branch}
   \addplot[black,dashed,domain=1.154701:5,samples=200]
   {-2*pi*x^2};
   \addlegendentry{Schwarzschild limit}
   \addplot[gray,densely dotted] coordinates
   {(1.1547005,-220) (1.1547005,220)};
   \addplot[red!70!black,dashdotted] coordinates
   {(1.7598771,-220) (1.7598771,220)};
  \end{axis}
 \end{tikzpicture}
 \caption{Exact heat capacity at fixed $\alpha$. The near-extremal branch
 (blue) has positive heat capacity. At $x_c$ (dash-dotted line) the
 temperature reaches its maximum and the heat capacity diverges; the
 large-black-hole branch (red) has negative heat capacity. This
 stability-transition feature was previously demonstrated in
 Ref.~\cite{ShiZhangMa2024}.}
 \label{fig:heat-capacity}
\end{figure}

At fixed $\alpha$,
\begin{equation}
 C_\alpha=\left(\frac{\partial M}{\partial T}\right)_\alpha
 =\frac{\alpha}{G}\,c(x),
\end{equation}
where direct differentiation of Eqs.~\eqref{eq:mass} and
\eqref{eq:theta} yields
\begin{equation}
 c(x)=
 \frac{2\pi x^3\left[xq+x^2-2\right]}
 {x(4-x^2)+q(2-x^2)}.                                            \label{eq:heat-capacity}
\end{equation}
The pole is determined by
\begin{equation}
 x(4-x^2)+\sqrt{x^2-1}(2-x^2)=0,
\end{equation}
whose physical solution is
\begin{equation}
 x_c=\sqrt{\frac{4+2\sqrt7}{3}}
 \simeq1.759877,\qquad
 \theta_c=\theta(x_c)\simeq0.0320366.                             \label{eq:critical}
\end{equation}
Thus $C_\alpha>0$ for $x_e<x<x_c$, it diverges at $x_c$, and
$C_\alpha<0$ for $x>x_c$. The large-$x$ limit,
$C_\alpha\sim-2\pi r_h^2/G$, agrees with Schwarzschild. Figure
\ref{fig:heat-capacity} is generated from the exact
Eq.~\eqref{eq:heat-capacity}; it replaces a polynomial illustrative curve
that did not encode the thermodynamics. The existence of the additional
heat-capacity transition was established in the related qOS analysis of
Shi, Zhang, and Ma \cite{ShiZhangMa2024}; our figure reproduces and
analytically locates it in the present four-dimensional parametrization.

\section{Canonical ensemble and saddle stability}
\label{sec:canonical}

\subsection{Definition and regulator}

For a system in contact with a heat bath at inverse temperature
$\beta=T_b^{-1}$, the canonical ensemble is defined by
\begin{equation}
 \rho_\beta=\frac{\ee^{-\beta H}}{Z(\beta)},\qquad
 Z(\beta)=\Tr\,\ee^{-\beta H},\qquad
 F(\beta)=-\beta^{-1}\ln Z(\beta).                               \label{eq:canonical-def}
\end{equation}
In Euclidean gravity the same object is evaluated semiclassically by a
functional integral over geometries periodic in Euclidean time,
\begin{equation}
 Z_{\mathcal R}(\beta,\alpha)
 =\int_{\mathcal R}\mathcal{D}g\,\ee^{-I_E[g]}
 \simeq\sum_s \ee^{-I_E[g_s]},
 \label{eq:path-integral}
\end{equation}
where $\mathcal R$ denotes boundary conditions or another regulator. The
subscript is essential: for an isolated asymptotically flat black hole,
$S(M)\sim4\pi GM^2$, and the formal density-of-states integral
\begin{equation}
 Z(\beta)\sim\int^\infty \dd M\,
 \exp\!\left[S(M)-\beta M\right]                                 \label{eq:divergent-Z}
\end{equation}
diverges. A finite cavity, for example, supplies a boundary temperature and
a finite energy range \cite{York1986}. The calculations below classify
local saddles using the asymptotic energy $M$; they do not replace the
boundary data required for a globally normalizable partition function.

\subsection{Off-shell action}

Define a dimensionless entropy and inverse bath temperature by
\begin{align}
 \sigma(x)&=\frac{G}{\alpha}S(x)
 =\pi\left[
 xq+\ln(x+q)-\frac{2}{3}-\frac{1}{2}\ln3
 \right],                                                        \label{eq:sigma}\\
 b&=\frac{\beta}{\sqrt{\alpha}}.
\end{align}
The reduced off-shell canonical action is
\begin{equation}
 \beta M-S=\frac{\alpha}{G}\,\mathcal{J}_b(x),\qquad
 \mathcal{J}_b(x)=b\,\mu(x)-\sigma(x).                            \label{eq:off-shell}
\end{equation}
The first law is equivalently
\begin{equation}
 \sigma'(x)=\frac{\mu'(x)}{\theta(x)}.                            \label{eq:first-law-dimless}
\end{equation}
Consequently,
\begin{equation}
 \mathcal{J}'_b(x_s)=0
 \quad\Longleftrightarrow\quad
 b=\frac{1}{\theta(x_s)}
 \quad\Longleftrightarrow\quad
 T_b=T(x_s).                                                      \label{eq:saddle}
\end{equation}
At a saddle,
\begin{equation}
 \mathcal{J}''_b(x_s)
 =\frac{\mu'(x_s)\theta'(x_s)}{\theta(x_s)^2}
 =\frac{\mu'(x_s)^2}{c(x_s)\theta(x_s)^2}.                       \label{eq:hessian}
\end{equation}
For $x_s>x_e$, the canonical saddle is therefore a local minimum precisely
when $C_\alpha>0$.

For every $0<T_b<T_c$, Eq.~\eqref{eq:saddle} has two solutions. The
near-extremal solution $x_e<x_s<x_c$ has $C_\alpha>0$ and is locally stable;
the larger solution $x_s>x_c$ has $C_\alpha<0$ and is unstable. They merge
at $T_b=T_c$, where the Gaussian saddle approximation becomes singular.
There is no black-hole saddle in this asymptotic parametrization for
$T_b>T_c$. This is a local stability statement. Comparing global free
energies with hot space or another phase requires a specified regulator,
boundary action, and common entropy normalization.

\section{Attribution, interpretation, and limitations}
\label{sec:scope}

To make the division between prior input and the analysis performed here
unambiguous, Table~\ref{tab:attribution} summarizes the status of the main
ingredients.

\begin{table}[htbp]
 \centering
 \caption{Provenance and role of the principal ingredients.}
 \label{tab:attribution}
 \begin{tabular}{p{0.27\textwidth}p{0.31\textwidth}p{0.34\textwidth}}
  \toprule
  Ingredient & Status & Role in this manuscript \\
  \midrule
  qOS exterior metric &
  Derived previously in Ref.~\cite{LewandowskiEtAl2023} &
  Adopted as the fixed geometric input; no new solution is claimed. \\
  Scalar wave equation &
  Analyzed for qOS-related metrics in
  Refs.~\cite{LinZhang2024,ShiZhangMa2024,OuZhang2025} &
  Restated only to fix notation and attribution. \\
  Temperature, logarithmic entropy, and heat-capacity transition &
  Established in the qOS literature
  \cite{LinZhang2024,ShiZhangMa2024,OuZhang2025,JiangLinZhang2026} &
  Reproduced exactly in one four-dimensional parametrization. \\
  Reduced canonical action and two-saddle map &
  Calculation organized in Sec.~\ref{sec:canonical} &
  Makes the local canonical minimum/maximum structure explicit; no
  priority claim is made for equivalent thermodynamic identities. \\
  \bottomrule
 \end{tabular}
\end{table}

Three limitations are material. First, Eqs.~\eqref{eq:metric} and
\eqref{eq:lapse} describe an effective exterior in a specific matching
construction, not a complete quantum geometry. Second, $\alpha$ is held
fixed. Promoting it to a thermodynamic variable would add a conjugate work
term and define a different ensemble. Third, the divergence of
$C_\alpha$ is a local stability transition. It is sometimes called an
additional black-hole phase transition in the literature
\cite{ShiZhangMa2024}, but a global phase transition can be asserted only
after the boundary-regulated free energies of all competing saddles have
been compared.

\section{Conclusion}
\label{sec:conclusion}

We have revised the analysis around a precise and limited objective:
canonical stability of an already known qOS black-hole exterior. The LQG
area formula is identified correctly as a spectrum on spin-network states,
and an unnecessary schematic volume formula has been removed. The adopted
metric is properly attributed, the dimension of $\alpha$ and the exterior
domain are stated, and no new regular solution is claimed.

At fixed $\alpha$, the exact horizon parametrization yields the Hawking
temperature, the first-law entropy with its logarithmic large-area term, and
the heat capacity. The corrected heat-capacity curve displays the known
divergence at
$x_c=\sqrt{(4+2\sqrt7)/3}$. An explicit canonical definition and off-shell
action then show that the near-extremal branch is a local minimum and the
large branch a local maximum. These conclusions are local and
regulator-dependent at the global level. A complete cavity calculation,
including quasilocal energy and a comparison with other phases, is the
appropriate next step.

\appendix
\section{Checks of the thermodynamic formulae}
\label{app:checks}

The horizon equation is quadratic in $m$,
\begin{equation}
 \alpha m^2-2r_h^3m+r_h^4=0.
\end{equation}
Choosing the Schwarzschild-connected root and rationalizing gives
Eq.~\eqref{eq:mass}. Its stationary point satisfies
$2\sqrt{r_h^2-\alpha}=r_h$, which yields Eq.~\eqref{eq:extremal}. Using the
horizon equation to eliminate $m$ from $f'(r_h)/(4\pi)$ gives
Eq.~\eqref{eq:temperature}.

For a direct first-law check, differentiating Eq.~\eqref{eq:entropy} gives
\begin{equation}
 \frac{\dd S}{\dd r_h}
 =\frac{2\pi r_h^2}{G\sqrt{r_h^2-\alpha}}.
\end{equation}
Together with Eqs.~\eqref{eq:mass} and \eqref{eq:temperature}, this identity
verifies $\dd M=T\,\dd S$. Differentiating $M(x)$ and $T(x)$ gives
Eq.~\eqref{eq:heat-capacity}. Squaring its pole equation is safe on the
physical interval $2<x^2<4$ and reduces it to
\begin{equation}
 3x^4-8x^2-4=0,
\end{equation}
whose positive physical root is Eq.~\eqref{eq:critical}.

\end{document}